\documentclass[twocolumn,showpacs,preprintnumbers,amsmath,amssymb,floatfix,prd]{revtex4}
\usepackage{graphicx}

\newcommand*{\comment}[1]{}

\newcommand*{\lr}[1]{ \left( #1 \right) }
\newcommand*{\lrs}[1]{ \left[ #1 \right] }

\providecommand*{\tr}{ {\rm Tr} \: }

\newcommand*{\expa}[1]{ \exp{\left( #1 \right)} }

\newcommand*{\zvar}[1]{ e^{\frac{2 \pi i}{N} \: #1  } }

\begin{document}
\sloppy
\preprint{ITEP-LAT/2007-07}

\title{Random walks of Wilson loops in the screening regime}
\author{P. V. Buividovich}
\email{buividovich@tut.by}
\affiliation{JIPNR, National Academy of Science, 220109 Belarus, Minsk, Acad. Krasin str. 99}
\author{M. I. Polikarpov}
\email{polykarp@itep.ru}
\affiliation{ITEP, 117218 Russia, Moscow, B. Cheremushkinskaya str. 25}
\date{July 10, 2007}
\begin{abstract}
 Dynamics of Wilson loops in pure Yang-Mills theories is analyzed in terms of random walks of the holonomies of the gauge field on the gauge group manifold. It is shown that such random walks should necessarily be free. The distribution of steps of these random walks is related to the spectrum of string tensions of the theory and to certain cumulants of Yang-Mills curvature tensor. It turns out that when colour charges are completely screened, the holonomies of the gauge field can change only by the elements of the group center, which indicates that in the screening regime confinement persists due to thin center vortices. Thick center vortices are also considered and the emergence of such stepwise changes in the limits of infinitely thin vortices and infinitely large loops is demonstrated.
\end{abstract}
\pacs{12.38.Aw; 05.40.Fb}
\maketitle

\section{Introduction}
\label{sec:Introduction}

 Wilson loops are the most popular order parameters for Yang-Mills theories without dynamical quarks. Wilson loop $W_{R}\lrs{C}$ is defined as the expectation value of the trace of the holonomy of the gauge field over the loop $C$: $W_{R}\lrs{C} = \langle \:  \tr_{R} \mathcal{P} \expa{i \int \limits_{C} dx^{\mu} A_{\mu}} \: \rangle$ \cite{Wilson:74}. Interaction potential $V_{R}\lr{r}$ between two static colour charges which transform under some irreducible representation $R$ of the gauge group can be measured using the Wilson loop $W_{R}\lrs{ C_{r \times t} }$, where $C_{r \times t}$ is a rectangular loop of size $r \times t$ with $t \rightarrow \infty$.  As the Wilson loop is in fact the amplitude of propagation of static colour charges along the loop $C$ \cite{Wilson:74}, it is related to the potential $V_{R}\lr{r}$ as $W_{R}\lrs{ C_{r \times t} } = \expa{- V_{R}\lr{r} \: t }$. It is known that in the confining phase of the theory static colour charges are connected by a thick chromoelectric string with constant tension $\sigma_{R}$, which gives rise to linear interaction potential $V_{R}\lr{r} = \sigma_{R} \: r$ at sufficiently large distances. Correspondingly, for sufficiently large loops Wilson loops $W_{R}\lrs{C}$ decay exponentially with the minimal area $S\lrs{C}$ of the surface spanned on the loop $C$, i.e. $W_{R}\lrs{C} = \expa{ - \sigma_{R} \: S\lrs{C}}$. Contribution of charge self-energies shows up in the dependence of $W_{R}\lrs{C}$ on the perimeter of the loop $C$.

 Thus confining properties of the theory are characterized by the whole spectrum of string tensions $\sigma_{R}$. The dependence of $\sigma_{R}$ on the representation of the gauge group gives some additional information on the properties of Yang-Mills theory. For instance, lattice simulations show that at intermediate distances ($0.2 - 1 \: {\rm fm}$ for $SU(3)$ gauge group) $\sigma_{R}$ is proportional to the eigenvalue $C_{2 \: R}$ of the second-order Casimir operator in the representation $R$ (Casimir scaling) \cite{Bali:00, Deldar:00}. Casimir scaling is naturally explained in the framework of stochastic vacuum models \cite{Dosch:02}. At very large distances Casimir scaling should be violated because of screening of colour charges by gluons \cite{Mack:78:1}. The reason is that when colour charges are separated by sufficiently large distance, it costs less energy to create a bound state of a colour charge and some number of gluons than to create confining string between bare colour charges. For instance, when $SU(N)$ colour charges are completely screened, string tension depends only on the $N$-ality of the representation \cite{Mack:78:1, Greensite:83, Douglas:95:1} and is proportional to the lowest eigenvalue of the second-order Casimir operator among all irreducible representations with the same $N$-ality.

 An interesting description of the spectrum of string tensions of Yang-Mills theories was proposed recently in \cite{Brzoska:05, Arcioni:05, Buividovich:06:2}, where all information about Wilson loops $W_{R}\lrs{C}$ in irreducible representations of the gauge group was encoded in a single function on the gauge group manifold, namely, the probability distribution $p\lrs{g; C}$ of the holonomies of the gauge field over the loop $C$. The holonomy $g\lrs{C} = \mathcal{P} \expa{i \int \limits_{C} dx^{\mu} A_{\mu} }$ is defined only modulo gauge transformations $g\lrs{C} \rightarrow h g\lrs{C} h^{-1}$, which implies that $p\lrs{g; C}$ should be invariant under such transformations, i.e. $p\lrs{g; C} = p\lrs{h g h^{-1}; C}$. Thus $p\lrs{g; C}$ should be defined as \cite{Brzoska:05, Arcioni:05, Buividovich:06:2}:
\begin{eqnarray}
\label{ProbDistrDef}
p\lrs{g; C} = \langle \delta_{c}\lr{g,g\lrs{C}} \rangle
\end{eqnarray}
where $\delta_{c}\lr{g,g'} = \sum \limits_{R} \bar{\chi}_{R}\lr{g} \chi_{R} \lr{g'}$ is the delta-function on the group classes and $\chi_{R}\lr{g}$ are the group characters. Using the character expansion of $\delta_{c}\lr{g,g'}$, one can immediately express $p\lrs{g; C}$ in terms of the Wilson loops $W_{R}\lrs{C}$:
\begin{eqnarray}
\label{ProbDistrWilsonLoops}
p\lrs{g; C} = \sum \limits_{R} \bar{\chi}_{R}\lr{g} W_{R}\lrs{C}
\end{eqnarray}
Conversely, all Wilson loops can be calculated if the probability distribution $p\lrs{g; C}$ is known: $W_{R}\lrs{C} = \int dg \: \chi_{R}\lr{g} \: p\lrs{g; C}$. It can be shown that $p\lrs{g; C}$ is the Wilson loop in the regular representation of the gauge group \cite{Buividovich:06:2, Buividovich:06:3}.

\begin{figure}
  \includegraphics[width=5cm]{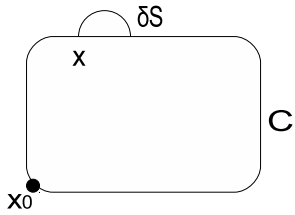}\\
  \caption{Small increment of the area of the minimal surface spanned on the loop $C$}
  \label{fig:loop}
\end{figure}

 An advantage of such description is that the evolution of $p\lrs{g; C}$ can be interpreted in terms of a random walk of the holonomy $g\lrs{C}$ on the gauge group manifold, the properties of this random walk being directly related to the spatial distribution of non-Abelian magnetic flux. In a particular configuration of gauge fields, the holonomy $g\lrs{C}$ moves along some path on the group manifold as the area of the loop is gradually increased. When one calculates expectation values of Wilson loops in quantum theory and sums over all field configurations, weighted sum over all such paths which start in the group identity and end in the element $g$ yields the probability distribution $p\lrs{g; C}$. In the case of pure Yang-Mills theory in Euclidean space-time all field configurations and, consequently, all such paths are summed with nonnegative weights, therefore such weighted sums over paths on the group manifold can be described as random walks on Lie groups \cite{Varopoulos:94, Guivarch:00:1}. The position of random walker corresponds then to the holonomy $g\lrs{C}$ modulo gauge transformations $g\lrs{C} \rightarrow h g\lrs{C} h^{-1}$. If the loop $C$ is slightly deformed in such a way that the area of the minimal surface increases from $S\lrs{C}$ to $S\lrs{C'} = S\lrs{C} + \delta S$ (see Fig. \ref{fig:loop}), the holonomy $g\lrs{C}$ changes as:
\begin{eqnarray}
\label{NAFluxAndSteps}
g\lrs{C'} = g\lrs{C} \lr{1 + i h\lr{x_{0},x} F_{\mu \nu}\lr{x} h\lr{x,x_{0}} \delta S^{\mu \nu} }
\end{eqnarray}
where $h\lr{x,y} = \mathcal{P} \expa{i \int \limits_{x}^{y} dx^{\mu} A_{\mu}}$ is the multiplicative integral of the gauge field over the segment of the loop $C$ bounded by the points $x$ and $y$. The point $x_{0}$ is the initial point used to calculate the holonomy $g\lrs{C}$: $g\lrs{C} = \mathcal{P} \expa{i \int \limits_{C; \: x_{0}}^{x_{0}} dx^{\mu} A_{\mu} }$. The combination $\tilde{F}_{\mu \nu}\lr{x} = h\lr{x_{0},x} F_{\mu \nu}\lr{x} h^{-1}\lr{x,x_{0}}$ is usually called the shifted curvature tensor \cite{Dosch:02}. Thus each step of the random walk of $g\lrs{C}$ corresponds to a small change of non-Abelian magnetic flux through the loop $C$.

 One of the basic observations of \cite{Brzoska:05, Arcioni:05, Buividovich:06:2} is that when Casimir scaling holds, the probability distribution $p\lrs{g; C}$ satisfies the free diffusion equation:
\begin{eqnarray}
\label{FreeDiffusionEquation}
\frac{d}{d S\lrs{C}} \: p\lrs{g; C} = \sigma_{0} \Delta p\lrs{g; C}
\end{eqnarray}
where $\Delta$ is the Laplace operator on the group manifold and $\sigma_{0}$ is some constant. Random walk which is described by (\ref{FreeDiffusionEquation}) is simply the Brownian motion on the group manifold, which consists of a large number of small statistically independent random steps. Correspondingly, the distribution of non-Abelian flux through the loops of intermediate sizes is almost random, which fits nicely in the model of stochastic vacuum \cite{Dosch:02}.

 Free diffusion equation (\ref{FreeDiffusionEquation}) and the corresponding random walk should be somehow modified in order to reproduce screening effects. In \cite{Brzoska:05, Arcioni:05} random walk in $Z_{N}$-symmetric potential was considered and it was shown that diffusion in such potential reproduces transition from Casimir scaling to complete screening due to mixing of different representations with equal $N$-alities. An advantage of such approach is that one can solve the diffusion equation in terms of path integral and obtain the effective action for Wilson loops. However, this approach is to a large extent phenomenological -- in particular, the choice of the potential which breaks $SU\lr{N}$ symmetry of the equation down to $Z_{N}$ is to a large extent arbitrary and affects only the transition from Casimir scaling to screening.

 In \cite{Buividovich:06:3} the most general form of the diffusion equation for $p\lrs{g; C}$ was obtained using the classical loop equations \cite{PolyakovGaugeStrings} and the cumulant expansion theorem \cite{VanKampenStochasticProcesses}. It turned out that this diffusion equation should have the form of the so-called  Kramers-Moyall cumulant expansion, which describes a free random walk with an arbitrary distribution of steps. Free random walk is understood here as a random walk with position-independent step distribution. Thus exact equation for $p\lrs{g; C}$ cannot include potential or drift terms.

 The aim of this paper is to consider the most general free random walk on the group manifold and to relate its properties to the spectrum of string tensions of Yang-Mills theory. Although in general it is not possible to find the effective action which describes a free random walk, the distribution of steps of the random walk of $g\lrs{C}$ has a more natural physical interpretation than an external potential in the diffusion equation. In particular, it will be shown that the distribution of steps of the random walk changes dramatically when Casimir scaling changes to screening. While the  free diffusion equation (\ref{FreeDiffusionEquation}) reflects stochasticity of Yang-Mills vacuum at small and intermediate distances, random walk which describes diffusion of Wilson loops in the screening regime can only consist of discrete jumps by the elements of the group center, which can be naturally explained in terms of thin center vortices \cite{tHooft:78, DelDebbio:97, Polikarpov:03:1}. It will be also demonstrated that in the case of center vortices with finite thickness the holonomy $g\lrs{C}$ can change by any group element, but the distribution of steps is still peaked near the elements of the group center, the widths of the peaks being proportional to vortex thickness and inversely proportional to the area $S\lrs{C}$ of the minimal surface spanned on the loop.

 The structure of the paper is the following: in the section \ref{sec:GeneralRandomWalk} a general case of a free random walk of $g\lrs{C}$ on the group manifold is considered. The distribution of steps of such random walk is expressed in terms of the spectrum of string tensions of the theory and in terms of the cumulants of the shifted curvature tensor $\tilde{F}_{\mu \nu}$. In the section \ref{sec:CompleteScreening} the random walk of $g\lrs{C}$ in the screening regime is investigated and it is shown that in the case of complete screening the holonomy $g\lrs{C}$ is only allowed to change by the elements of the group center. In the section \ref{sec:ThinCenterVortices} such discrete jumps are interpreted in terms of thin center vortices. Vortices of finite thickness are considered in the section \ref{sec:ThickCenterVortices}, where it is shown how such jumps by the elements of the group center arise in the limits of infinitely thin vortices or infinitely large loops.

\section{Free random walks on the group manifold}
\label{sec:GeneralRandomWalk}

 In \cite{Buividovich:06:3} it was shown that the most general form of the diffusion equation for the probability distribution of the holonomies $g\lrs{C}$ on the gauge group manifold is:
\begin{eqnarray}
\label{KramersMoyall}
\frac{d}{d S\lrs{C}} \: p\lrs{g; C} = \sum \limits_{k=2}^{\infty} \eta^{a_{1} \ldots  a_{k}}\lrs{C} \nabla_{a_{1}} \ldots \nabla_{a_{k}} \: p\lrs{g; C}
\end{eqnarray}
where $\eta^{a_{1} \ldots  a_{k}}\lrs{C}$ are some coefficients which can in general depend on the loop $C$ and $\nabla_{a}$ are the generators of left shifts on the group manifold. The action of $\nabla_{a}$ is defined by the identity $\lr{1 + \epsilon^{a} \nabla_{a}} f\lr{g} = f\lr{\lr{1 + i \epsilon^{a} T_{a}} g}$, where $\epsilon^{a}$ are arbitrary infinitely small parameters, $T_{a}$ are the generators of the gauge group and $f\lr{g}$ is an arbitrary function on the group manifold. The term with first-order derivative is prohibited in (\ref{KramersMoyall}) by gauge invariance. It was shown in \cite{Buividovich:06:3} that the coefficients $\eta^{a_{1} \ldots  a_{k}}\lrs{C}$ are directly related to the cumulants of the shifted curvature tensor $\tilde{F}^{a}_{\mu \nu}$, which are the basic objects in the method of field correlators \cite{Dosch:02}. This relation can be formally written as \cite{Buividovich:06:3}:
\begin{widetext}
\begin{eqnarray}
\label{DiffusionCoefficients}
\eta^{a_{1} \ldots  a_{k}}\lrs{C} =  \frac{ \lr{-1}^{k}}{k!} \frac{d}{d S\lrs{C}}
\int \limits_{S\lrs{C}} \ldots \int \limits_{S\lrs{C}}
d S^{\mu_{1} \nu_{1}}\lr{x_{1}} \ldots d S^{\mu_{k} \nu_{k}}\lr{x_{k}}
\langle \langle \tilde{F}^{a_{1}}_{\mu_{1} \nu_{1}}\lr{x_{1}} \ldots  \tilde{F}^{a_{k}}_{\mu_{k} \nu_{k}} \lr{x_{k}} \rangle \rangle
\end{eqnarray}
where the integration is performed over the surface of the minimal area $S\lrs{C}$ spanned on the loop $C$. It should be noted that for any point $x$ the shifted curvature tensor $\tilde{F}^{a}_{\mu \nu}\lr{x}$ transforms under gauge transformations as the curvature tensor in some fixed point $x_{0}$, therefore the cumulants of $\tilde{F}^{a}_{\mu \nu}\lr{x}$ are well-defined when all $x_{1}, \ldots, x_{k}$ are different. Gauge invariance implies only that the cumulants of $\tilde{F}^{a}_{\mu \nu}\lr{x}$ and the coefficients $\eta^{a_{1} \ldots  a_{k}}\lrs{C}$ should be colour singlets. In particular, all $\eta^{a_{1} \ldots  a_{k}}\lrs{C}$ with odd $k$ should vanish.

 The equation (\ref{KramersMoyall}) is a general equation which describes an arbitrary free random walk on the group manifold, i.e. a random walk with position-independent distribution of steps. If $P\lrs{g'; C} \delta S$ is the probability that $g\lrs{C}$ changes to $g\lrs{C'} = g' g\lrs{C}$ as the area of the surface spanned on the loop $C$ increases from $S\lrs{C}$ to $S\lrs{C'} = S\lrs{C} + \delta S$, the evolution of the probability distribution $p\lrs{g; C}$ is described by the following equation:
\begin{eqnarray}
\label{DiffEqIntegralGeneral}
\frac{d}{d S\lrs{C}} \: p\lrs{g; C}
 =
\int dg' P\lrs{g' g^{-1}; C} p\lrs{g'; C} - p\lrs{g; C} \int dg' P\lrs{g'; C}
 = \nonumber \\ =
\int dg' \lr{ P\lrs{g' g^{-1}; C} - \delta\lr{g',g} \int dg'' P\lrs{g''; C} } p\lrs{g'; C}
\end{eqnarray}
where $\delta\lr{g',g}$ is the delta-function on the group manifold. Although the distribution of steps in  (\ref{DiffEqIntegralGeneral}) can in general depend on the loop $C$, this dependence can be neglected if the Wilson area law holds exactly and if the spectrum of string tensions $\sigma_{R}$ is known. Using the definition (\ref{ProbDistrDef}) the following integral equation can be obtained for $p\lrs{g; C}$:
\begin{eqnarray}
\label{DiffEqIntegral}
\frac{d}{d S \lrs{C}} \: p\lrs{g; C} = - \sum \limits_{R} \bar{\chi}_{R}\lr{g} \sigma_{R} W_{R}\lrs{ C }
= - \sum \limits_{R} \bar{\chi}_{R}\lr{g} \sigma \lr{R} \int dg' p\lrs{g'; C} \chi_{R}\lr{g'}
= \nonumber \\ =
 - \int dh \int dg' \sum \limits_{R} d_{R} \bar{\chi}_{R}\lr{g h g'^{-1} h^{-1}} \sigma \lr{R}  p\lrs{g'; C}
 =
\int dg' \lr{ - \sum \limits_{R} d_{R} \bar{\chi}_{R}\lr{g g'^{-1}} \sigma \lr{R} } p\lrs{g'; C}
\end{eqnarray}
\end{widetext}
where the identity $\int dh \chi_{R}\lr{g h f h^{-1}} = d_{R}^{-1} \chi_{R}\lr{g} \chi_{R} \lr{f}$ and the invariance of $p \lrs{g; C}$ w.r.t. gauge transformations $g\lrs{C} \rightarrow h g\lrs{C} h^{-1}$ were used. Comparing the equations (\ref{DiffEqIntegral}) and (\ref{DiffEqIntegralGeneral}), one can finally express the step distribution $P\lrs{g; C}$ in terms of string tensions $\sigma_{R}$:
\begin{eqnarray}
\label{StepDistrFromSpectrum}
P\lrs{g; C} =  - \sum \limits_{R} d_{R} \: \sigma_{R} \: \bar{\chi}_{R}\lr{g}
\end{eqnarray}
If string tensions depend on the distance between colour charges, for instance, because of screening, $P\lrs{g; C}$ should also depend on the loop $C$. This dependence will be investigated in the section \ref{sec:ThickCenterVortices} using the model of thick center vortices.  It is also interesting to note that the positivity of  $P\lrs{g; C}$ for $g \neq 1$ is some constraint on possible spectrum of string tensions of the theory, which follows from the positivity of path integral weight for pure Yang-Mills theory in Euclidean space-time.

 The equation (\ref{KramersMoyall}) is simply the gradient expansion of the general equation (\ref{DiffEqIntegralGeneral}). Indeed, $p\lr{g'}$ in (\ref{DiffEqIntegralGeneral}) can be replaced by $p\lrs{g'g^{-1} g; C} = \expa{- \xi^{a}\lr{g g'{}^{-1}} \nabla_{a} } p\lrs{g; C}$, where the function $\xi^{a}\lr{g}$ is defined by the following identity:
\begin{eqnarray}
\label{GroupLogarithm}
 \expa{i \xi^{a}\lr{g} T_{a} } = g
\end{eqnarray}
 Expanding the exponential in powers of $\xi^{a}\lr{g g'{}^{-1}}$, one can express the coefficients $\eta^{a_{1} \ldots  a_{k}}\lrs{C}$ in terms of the step distribution $P\lrs{g; C}$:
\begin{eqnarray}
\label{CumulantsFromStepDistr}
\eta^{a_{1} \ldots  a_{k}}\lrs{C} = \frac{\lr{-1}^{k}}{k!} \int dg \: P\lrs{g; C} \: \xi^{a_{1}}\lr{g} \ldots \xi^{a_{k}}\lr{g}
\end{eqnarray}
As all $\eta^{a_{1} \ldots  a_{k}}\lrs{C}$ are colour singlets, the distribution of steps should satisfy $P\lrs{h g h^{-1}; C} = P\lrs{g; C}$.

 For example, if Casimir scaling holds, $\sigma_{R} = \sigma_{0} C_{2 \: R}$, $P\lrs{g; C} = \sigma_{0} \Delta \delta\lr{g,1}$ and only the coefficient $\eta^{a b}\lrs{C} = \sigma_{0} \delta_{a b}$ is not equal to zero. Correspondingly, only the second-order cumulant of the shifted curvature tensor is nonzero, which corresponds to the limit of gaussian-dominated stochastic vacuum \cite{Dosch:02}.

\section{Complete screening and discrete jumps on the group manifold}
\label{sec:CompleteScreening}

 In the case of complete screening string tensions $\sigma_{R}$ depend only on the $N$-ality of the representation $R$: $\sigma_{R} = \sigma\lr{\Lambda_{R}}$. $N$-ality $\Lambda_{R}$ of the representation $R$ characterizes its transformation properties w.r.t. the elements of the group center $Z_{N} = \{ \zvar{k} \}$, $k = 0, 1, \ldots, N-1$: $\chi_{R} \lr{g z} = \chi_{R} \lr{g} z^{\Lambda_{R}}$, $z \in Z_{N}$. A simple calculation shows that if $\sigma_{R} = \sigma\lr{\Lambda_{R}}$, $P\lrs{g; C}$ is given by a finite sum over the elements of the group center:
\begin{eqnarray}
\label{ScrDistr}
P\lrs{g; C} = \sum \limits_{z \in Z_{N}} \eta\lr{z} \delta\lr{g,z}
\end{eqnarray}
where $\eta \lr{z}$ is minus the Fourier transform of the spectrum of string tensions w.r.t. the group center:
\begin{eqnarray}
\label{StrTensFourier}
\eta \lr{z} = - N^{-1} \sum \limits_{\Lambda = 0}^{N - 1} \sigma\lr{\Lambda} z^{\Lambda}
\end{eqnarray}
Thus when colour charges are completely screened, the evolution of the probability distribution $p\lrs{g; C}$ is described by the following equation:
\begin{eqnarray}
\label{ScrDiffusionEquation}
\frac{d}{d S \lrs{C}} \: p \lrs{g; C} = \sum \limits_{z \in Z_{N}} \eta\lr{z} \: p\lrs{z g; C}
\end{eqnarray}

 This equation implies that the random walk of $g\lrs{C}$ consists of jumps by the elements of the group center only. $\eta\lr{z}$ with $z \neq 1$ is the probability of a jump by $z$ per unit area. As the string tension between charges with zero $N$-ality should vanish due to complete screening by gluons \cite{Mack:78:1}, $\sum \limits_{z \in Z_{N}} \eta \lr{z} = 0$, which is simply the conservation of probability flow in the language of random walks.

 Thus the distribution of steps of the random walk of $g\lrs{C}$ is completely different in Casimir scaling and in screening regimes. When Casimir scaling holds, random walk of $g\lrs{C}$ consists of a large number of statistically independent small steps, which indicates that non-Abelian magnetic flux which penetrates the loop $C$ fluctuates randomly. On the other hand, the equation (\ref{ScrDiffusionEquation}) implies that the cumulants in (\ref{DiffusionCoefficients}) are saturated by some singular field configurations.

\section{Thin center vortices}
\label{sec:ThinCenterVortices}

 It is straightforward to guess what is the structure of singular field configurations which correspond to discrete jumps described by (\ref{ScrDiffusionEquation}). By definition the holonomy $g\lrs{C}$ changes by the element of the group center when the loop $C$ is crossed by center vortex \cite{tHooft:78}. As follows from equation (\ref{ScrDiffusionEquation}), $g\lrs{C}$ changes stepwise, which can only be explained if the vortices are mathematically thin surfaces which are distributed in space with some finite density. Correspondingly, $\eta\lr{z} \delta S$ is the probability to cross thin center vortex which carries magnetic flux $z$ as the area of the loop is increased from $S\lrs{C}$ to $S\lrs{C} + \delta S$.

 The statement about the singular field configurations can be formulated more precisely using the equations (\ref{DiffusionCoefficients}) and (\ref{CumulantsFromStepDistr}), which relate the step distribution $P\lrs{g; C}$, the coefficients $\eta^{a_{1} \ldots a_{k}}\lrs{C}$ and the cumulants of the shifted curvature tensor of Yang-Mills fields. The coefficients $\eta^{a_{1} \ldots a_{k}}\lrs{C}$ can be calculated if the field configurations which contribute to the cumulants in (\ref{DiffusionCoefficients}) are known. On the other hand, these coefficients can be found from the step distribution $P\lrs{g; C}$. In this section it will be shown that the coefficients $\eta^{a_{1} \ldots a_{k}}\lrs{C}$ obtained from the distribution (\ref{ScrDistr}) agree with the result obtained from (\ref{DiffusionCoefficients}) if center vortices are infinitely thin non-interacting random surfaces. This analysis will be extended in the next section, where center vortices of finite thickness will be considered and the emergence of the step distribution (\ref{ScrDistr}) in the limit of infinitely thin vortices will be demonstrated.

 In order to calculate the coefficients $\eta^{a_{1} \ldots a_{k}}\lrs{C}$ for the step distribution (\ref{ScrDistr}), one should calculate the integrals of the form $\int dg \delta \lr{g,z} \xi^{a_{1}}\lr{g} \ldots \xi^{a_{k}}\lr{g}$. Such integrals, however, can not be calculated by direct substitution $g \rightarrow z$, as for the elements of the group center the choice of $\xi^{a}\lr{g}$ is not unique. In fact, the points $g = z$ with $z \neq 1$ are the only points on the group  manifold where the function $\xi^{a}\lr{g}$ is not uniquely defined. Indeed, as $\expa{i \xi^{a}\lr{z} T_{a} } = z$, for any group element $h$ one has $h \expa{i \xi^{a}\lr{z} T_{a} } h^{-1} = \expa{i \xi^{a}\lr{z} h T_{a} h^{-1} } = \expa{i \xi^{b}\lr{z} O^{a}_{b}\lr{h} T_{a} } = h z h^{-1} = z$ (here $O^{a}_{b}\lr{h}$ are the matrices of the adjoint representation of the gauge group). The integrals $\int dg \delta \lr{g,z} \xi^{a_{1}}\lr{g} \ldots \xi^{a_{k}}\lr{g}$ can nevertheless be calculated if the delta-functions in (\ref{ScrDistr}) are defined as the limits of some smooth functions, which immediately yields $\int dg \delta \lr{g,z} \xi^{a_{1}}\lr{g} \ldots \xi^{a_{k}}\lr{g} = \int dh \: O_{b_{1}}^{a_{1}}\lr{h} \xi^{b_{1}}\lr{z} \ldots O_{b_{k}}^{a_{k}}\lr{h} \xi^{b_{k}}\lr{z}$. It is convenient to introduce the notation $\langle \xi^{a_{1}}\lr{z} \ldots \xi^{a_{k}}\lr{z} \rangle_{h} = \int dh \: O_{b_{1}}^{a_{1}}\lr{h} \xi^{b_{1}}\lr{z} \ldots O_{b_{k}}^{a_{k}}\lr{h} \xi^{b_{k}}\lr{z} $, so that the coefficients $\eta^{a_{1} \ldots a_{k}}\lrs{C}$ which correspond to the step distribution (\ref{ScrDistr}) can be written as:
\begin{eqnarray}
\label{ScrCumulants}
\eta^{a_{1} \ldots a_{k}}\lrs{C}
=
\frac{\lr{-1}^{k}}{k!} \sum \limits_{z \in Z_{N}} \eta\lr{z}
\langle \: \xi^{a_{1}}\lr{z} \ldots \xi^{a_{k}}\lr{z} \: \rangle_{h}
\end{eqnarray}
The equations (\ref{DiffusionCoefficients}) and (\ref{ScrCumulants}) imply that in this case the integrals of the cumulants of the shifted curvature tensor should have the following form:
\begin{widetext}
\begin{eqnarray}
\label{ScrFieldTensors}
\int \limits_{S\lrs{C}} \ldots \int \limits_{S\lrs{C}}
d S^{\mu_{1} \nu_{1}}\lr{x_{1}} \ldots d S^{\mu_{k} \nu_{k}}\lr{x_{k}}
\langle \langle \:  \tilde{F}^{a_{1}}_{\mu_{1} \nu_{1}}\lr{x_{1}} \ldots  \tilde{F}^{a_{k}}_{\mu_{k} \nu_{k}} \lr{x_{k}} \: \rangle \rangle
 = 
\sum \limits_{z \in Z_{N}} \eta\lr{z} \: S\lrs{C} \: \langle \: \xi^{a_{1}}\lr{z} \ldots \xi^{a_{k}}\lr{z} \: \rangle_{h}
\end{eqnarray}

 On the other hand, the integrals in (\ref{ScrFieldTensors}) can be calculated using the exact expression for the field strength of thin center vortex \cite{Engelhardt:00:2}:
\begin{eqnarray}
\label{VortexField}
F^{a}_{\mu \nu}\lr{x} = \xi^{a}\lr{z} \: \epsilon_{\mu \nu \alpha \beta} \: \int \limits_{\Sigma} d \Sigma^{\alpha \beta}\lr{y} \delta^{4}\lr{x - y}
\end{eqnarray}
where $\Sigma$ is the vortex surface. The choice of $\xi^{a}\lr{z}$ in (\ref{VortexField}) is also not unique, and it is usually assumed that physical observables should be averaged over all possible $\xi^{a}\lr{z}$ \cite{Ambjorn:80, Nielsen:79:1}. It is convenient now to introduce the linking number of the vortex surface $\Sigma$ and the loop $C$:
\begin{eqnarray}
\label{LinkingNumberDef}
\mathbb{L}\lrs{C; \Sigma} = \epsilon_{\alpha \beta \mu \nu} \: \int \limits_{S\lrs{C}} dS^{\mu \nu}\lr{x} \int \limits_{\Sigma}  d \Sigma^{\alpha \beta}\lr{y} \delta^{4}\lr{x - y}
\end{eqnarray}
The linking number $\mathbb{L}\lrs{C; \Sigma}$ counts how many times the surface $\Sigma$ winds around the loop $C$. Using the expressions (\ref{VortexField}), (\ref{LinkingNumberDef}) and averaging over all possible vortex configurations and over all $\xi^{a}\lr{z}$, one can rewrite the integrals in (\ref{ScrFieldTensors}) as:
\begin{eqnarray}
\label{VortexFieldCumulants}
\int \limits_{S\lrs{C}} \ldots \int \limits_{S\lrs{C}}
d S^{\mu_{1} \nu_{1}}\lr{x_{1}} \ldots d S^{\mu_{k} \nu_{k}}\lr{x_{k}}
\langle \langle \:  \tilde{F}^{a_{1}}_{\mu_{1} \nu_{1}}\lr{x_{1}} \ldots  \tilde{F}^{a_{k}}_{\mu_{k} \nu_{k}} \lr{x_{k}} \: \rangle \rangle
 =
\sum \limits_{z \in Z_{N}}
\langle \langle \: \mathbb{L}^{k}\lrs{C; \Sigma_{z}} \: \xi^{a_{1}}\lr{z} \ldots \xi^{a_{k}}\lr{z} \: \rangle \rangle
\end{eqnarray}
where $\Sigma_{z}$ denotes the union of the surfaces of all vortices which carry magnetic flux $z$. As only the cumulants of even order are different from zero, all indices $a_{1}, \ldots, a_{k}$ can be pairwise contracted in both (\ref{VortexFieldCumulants}) and (\ref{ScrFieldTensors}). As for all possible $\xi^{a}\lr{z}$ the squares  $\xi^{a}\lr{z} \xi^{a}\lr{z}$ are equal, this gives equal constant factors in front of both expressions. After that direct comparison of (\ref{VortexFieldCumulants}) and (\ref{ScrFieldTensors}) shows that all even-order cumulants of the linking number $\mathbb{L}\lrs{C; \Sigma_{z}}$ are equal: $ \langle \langle \: \mathbb{L}^{2 k}\lrs{C; \Sigma_{z}} \: \rangle \rangle = \eta\lr{z} S\lrs{C}$, $ \langle \langle \: \mathbb{L}^{2 k + 1} \lrs{C; \Sigma_{z}} \: \rangle \rangle = 0$. This property is enough to recover the probability distribution of the linking number $\mathbb{L}\lrs{C; \Sigma_{z}}$ -- it is distributed as the difference of two Poisson-distributed positive integer numbers $k_{+}$ and $k_{-}$ with mean values $\eta\lr{z} S\lrs{C} / 2$ each. The numbers $k_{+}$ and $k_{-}$ can be readily interpreted as the numbers of vortices which wind around $C$ leftwards and rightwards. Explicit expression for the probability distribution of $\mathbb{L}\lrs{C; \Sigma_{z}}$ is:
\begin{eqnarray}
\label{PoissonDiffDistr}
p\lr{ \mathbb{L}\lrs{C; \Sigma_{z}} } =
\expa{ - \eta\lr{z} \: S\lrs{C} } \sum \limits_{ k_{+} - k_{-} = \mathbb{L}\lrs{C; \Sigma_{z}} }
\frac{\lr{\eta\lr{z} \: S\lrs{C}/2}^{k_{+} + k_{-}}}{ k_{+}! \: k_{-}! }
\end{eqnarray}
\end{widetext}

 The fact that $k_{+}$ and $k_{-}$ are distributed by Poisson implies that thin vortices do not interact, since all intersections of vortex surfaces $\Sigma_{z}$ and the loop $C$ are statistically independent events. Recent results of lattice simulations indicate that center vortices are indeed mathematically thin random surfaces which are distributed in space with finite density $\rho_{vort} \approx 24 \: fm^{-2}$ \cite{Polikarpov:03:1}. A typical vortex configuration on the lattice consists of a single percolating vortex which stretches through the whole physical space plus a large number of small vortices, whose sizes do not exceed several lattice spacings \cite{Polikarpov:03:1, Polikarpov:03:2}. As such small vortices can only lead to perimeter-dependent effects, the existence of the percolating vortex is crucial for the area dependence of the expectation values $\langle k_{+} \rangle = \langle k_{-} \rangle = \eta\lr{z} S\lrs{C}/2$. The results obtained in \cite{Buividovich:07:3} also give some preliminary indications that the effective interaction between thin vortices is rather small. Thus the picture of thin non-interacting center vortices may be a good approximation for the low-energy dynamics of pure Yang-Mills theories.

\section{Thick center vortices}
\label{sec:ThickCenterVortices}

 The analysis presented in the previous section is based on the assumption that thin center vortices are the only field configurations which contribute to the cumulants of the shifted curvature tensor in (\ref{DiffusionCoefficients}), which is justified if the equation (\ref{ScrDiffusionEquation}) is exact for loops of finite size. However, since complete screening is only an asymptotic law which holds in the limit of very large loops with $S\lrs{C} \rightarrow \infty$, (\ref{ScrDiffusionEquation}) could also be the limiting case of some more general diffusion equation which takes into account the contribution of other field configurations. In order to derive such more general equation, one should make some model-dependent assumptions on the form of the cumulants of the shifted curvature tensor, so that the coefficients $\eta^{a_{1} \ldots a_{k}}\lrs{C}$ can be calculated explicitly. In this section the diffusion equation (\ref{DiffEqIntegralGeneral}) will be analyzed under the assumption that the dominating field configurations in the vacuum of Yang-Mills theory are thick center vortices, which roughly corresponds to the picture of "spaghetti vacuum" \cite{Ambjorn:80, Nielsen:79:1}. Although the assumptions and approximations made below by no means capture all properties of the theory, they are enough to demonstrate how the equation (\ref{ScrDiffusionEquation}) emerges in the limit of infinitely thin vortices or in the limit of very large loops. Vortices of finite thickness can be considered as an effective description of the interference between thin center vortices which were observed in lattice simulations \cite{Polikarpov:03:1} and other field configurations, as far as such picture of the vacuum of Yang-Mills theory correctly reproduces the spectrum of string tensions.

\begin{figure}
  \includegraphics[width=7cm]{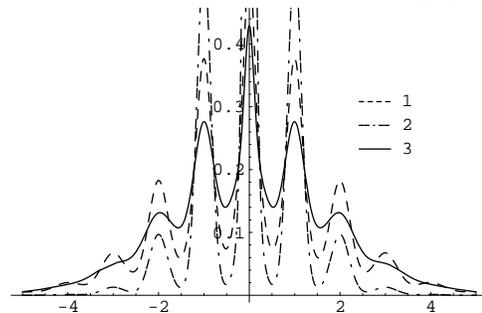}\\
  \caption{The probability distribution of $\tilde{\mathbb{L}}\lrs{C; \Sigma_{z}}$ for different values of $\eta\lr{z} S\lrs{C}$ and $\epsilon\lrs{C}$: $\eta\lr{z} S\lrs{C} = 3$, $\epsilon\lrs{C} = 0.05$ for plot 1, $\eta\lr{z} S\lrs{C} = 1$, $\epsilon\lrs{C} = 0.05$ for plot 2, $\eta\lr{z} S\lrs{C} = 3$, $\epsilon\lrs{C} = 0.1$ for plot 3}
  \label{fig:distrib}
\end{figure}

 Field strength of a thick center vortex can be obtained by smoothing the delta-function in (\ref{VortexField}) \cite{Engelhardt:00:2}:
\begin{eqnarray}
\label{ThickVortexField}
F^{a}_{\mu \nu}\lr{x} = \xi^{a}\lr{z} \: \epsilon_{\mu \nu \alpha \beta} \: \int \limits_{\Sigma} d \Sigma^{\alpha \beta}\lr{y} f\lr{x - y}
\end{eqnarray}
where $f\lr{x-y}$ is some smooth function which decays at distances of order of the vortex thickness $l_{vort}$.

 Proceeding as in the previous section, one can calculate the integrals of the cumulants of the shifted curvature tensor for the field configurations (\ref{ThickVortexField}), assuming that vortex surfaces $\Sigma_{z}$ are random. However, for thick center vortices the linking number $\mathbb{L}\lrs{C; \Sigma}$ can not be rigorously defined. Its role is now played by the integral of the following form:
\begin{eqnarray}
\label{ModifiedLinking}
\tilde{\mathbb{L}}\lrs{C; \Sigma} = \epsilon_{\mu \nu \alpha \beta} \int \limits_{S\lrs{C}} dS^{\mu \nu}\lr{x} \int \limits_{\Sigma} d\Sigma^{\alpha \beta}\lr{y} f\lr{x - y}
\end{eqnarray}
Just as in (\ref{VortexFieldCumulants}), the cumulants of the shifted curvature tensor can be expressed in terms of $\tilde{\mathbb{L}}\lrs{C; \Sigma}$:
\begin{widetext}
\begin{eqnarray}
\label{ThickVortexFieldCumulants}
\int \limits_{S\lrs{C}} \ldots \int \limits_{S\lrs{C}}
d S^{\mu_{1} \nu_{1}}\lr{x_{1}} \ldots d S^{\mu_{k} \nu_{k}}\lr{x_{k}}
\langle \langle \:  \tilde{F}^{a_{1}}_{\mu_{1} \nu_{1}}\lr{x_{1}} \ldots  \tilde{F}^{a_{k}}_{\mu_{k} \nu_{k}} \lr{x_{k}} \: \rangle \rangle
 = 
\sum \limits_{z \in Z_{N}}
\langle \langle \: \tilde{\mathbb{L}}^{k}\lrs{C; \Sigma_{z}} \: \xi^{a_{1}}\lr{z} \ldots \xi^{a_{k}}\lr{z} \: \rangle \rangle
\end{eqnarray}
On the other hand, the integrals in (\ref{ThickVortexFieldCumulants}) can be related to the coefficients $\eta^{a_{1} \ldots a_{k}}\lrs{C}$ in (\ref{KramersMoyall}) using the equation (\ref{DiffusionCoefficients}). These coefficients can be also obtained from $P\lrs{g; C}$ using the equation (\ref{CumulantsFromStepDistr}). Comparing  (\ref{DiffusionCoefficients}), (\ref{ThickVortexFieldCumulants}) and (\ref{CumulantsFromStepDistr}), one can obtain an equation which relates the cumulants of $\tilde{\mathbb{L}}\lrs{C; \Sigma_{z}}$ and the averages of $\xi^{a}\lr{g}$:
\begin{eqnarray}
\label{GroupCumulantsVsLNDistr}
S\lrs{C} \: \int dg \: P\lrs{g; C} \: \xi^{a_{1}}\lr{g} \ldots \xi^{a_{k}}\lr{g}
 =
\sum \limits_{z \in Z_{N}} \langle \langle \: \tilde{\mathbb{L}}^{k}\lrs{C; \Sigma_{z}} \: \xi^{a_{1}}\lr{z} \ldots \xi^{a_{k}}\lr{z} \: \rangle \rangle
\end{eqnarray}

This relation can be used to find the step distribution $P\lrs{g; C}$ if the distribution of $\tilde{\mathbb{L}}\lrs{C; \Sigma_{z}}$ is known. $\tilde{\mathbb{L}}\lrs{C; \Sigma_{z}}$ is not in general integer-valued, but it takes integer values when all thick vortices which contribute to the integral (\ref{ModifiedLinking}) are completely inside the loop $C$. Deviations from integer values occur if some vortices are only partially within the loop. If the loop is pierced by $k_{+}$ right-winding vortices and $k_{-}$ left-winding vortices, the average number of such vortices can be estimated as $\delta \tilde{\mathbb{L}}\lrs{C; \Sigma_{z}} \sim \lr{ k_{+} + k_{-} } l_{vort} \: P\lrs{C} / S\lrs{C}$, where $P\lrs{C}$ is the perimeter of the loop $C$. For sufficiently large loops and sufficiently small vortex densities one can assume that $\delta \tilde{\mathbb{L}}\lrs{C; \Sigma_{z}}$ and $k_{+}$, $k_{-}$ are also distributed by Poisson, i.e. that the interaction between vortices can be neglected. As a nearest extension of the results obtained in the previous section, one can try to smear the original distribution of linking numbers (\ref{PoissonDiffDistr}) in such a way that the probability distribution $p\lr{ \tilde{\mathbb{L}}\lrs{C; \Sigma_{z}} }$ is still peaked around integer values, the widths of the peaks being proportional to $\delta \mathbb{L}\lrs{C; \Sigma_{z}}$. Without loss of generality the shape of the peaks can be approximated by a gaussian distribution. The resulting probability distribution is:
\begin{eqnarray}
\label{PoissonDiffDistrSmeared}
p\lr{ \tilde{\mathbb{L}}\lrs{C; \Sigma_{z}} } = \expa{ - \eta\lr{z} \: S\lrs{C} } 
\sum \limits_{ k_{+}, k_{-} }
\frac{\lr{\eta\lr{z} \: S\lrs{C}/2}^{k_{+} + k_{-}}}{ k_{+}! \: k_{-}! } \: G\lr{\tilde{\mathbb{L}}\lrs{C; \Sigma_{z}} - \lr{k_{+} - k_{-}}, \lr{ k_{+} + k_{-} } \epsilon\lrs{C} }
\end{eqnarray}
where $\epsilon\lrs{C} = l_{vort} \: P\lrs{C}/S\lrs{C}$ is assumed to be small and $G\lr{ x, \sigma^{2} } = \lr{2 \pi \sigma^{2} }^{-1/2} \expa{ - x^{2}/\lr{2 \sigma^{2}} }$ is the gaussian distribution with dispersion $\sigma$. Such probability distribution for some values of $\eta\lr{z} S\lrs{C}$ and $\epsilon\lrs{C}$ is plotted on Fig. \ref{fig:distrib}.  It turns out that for the distribution (\ref{PoissonDiffDistrSmeared}) the generating function for the cumulants can be found explicitly:
\begin{eqnarray}
\label{PoissonSmearedConnectedGenFunc}
W\lr{ q, z } = \sum \limits_{k = 1}^{ + \infty} \frac{\lr{i q}^{k}}{k!} \: \langle \langle \: \tilde{\mathbb{L}}^{k}
\lrs{C; \Sigma_{z}} \: \rangle \rangle
 =
\eta\lr{z} \: S\lrs{C} \: \lr{ \cos q \expa{ - \epsilon\lrs{C} q^{2} } - 1}
\end{eqnarray}
For comparison, for the distribution (\ref{PoissonDiffDistr}) $W\lr{q,z} = \eta\lr{z} \: S\lrs{C} \: \lr{\cos q - 1}$.

For illustration, the step distribution $P\lrs{g; C}$ which solves (\ref{GroupCumulantsVsLNDistr}) will be found for $SU\lr{2}$ gauge group.  For $SU\lr{2}$ group $z = \pm 1$, $\xi^{2}\lr{1} = 0$ and $\xi^{2}\lr{-1} = \pi^{2}$ if the generators $T_{a}$ are the Pauli matrices. For the cumulants of even order the indices $a_{1}, \ldots, a_{2 k}$ can be again pairwise contracted. As $P\lrs{g; C}$ is the function on the group classes, it depends only on the absolute value of $\xi^{a}\lr{g}$. Taking into account that the Haar measure on $SU\lr{2}$ group classes is $dg = 2/\pi \sin^{2} \xi \: d \xi$, one can rewrite the relation (\ref{GroupCumulantsVsLNDistr}) as:
\begin{eqnarray}
\label{GroupCumulantsVsLNDistrContracted}
2/\pi S\lrs{C} \int \limits_{0}^{\pi} d\xi \: \sin^{2}\xi \: P\lrs{\xi; C} \lr{\xi / \pi }^{2 k}
 =
\langle \langle \: \tilde{\mathbb{L}}^{2 k}\lrs{C; \Sigma_{-1}} \: \rangle \rangle
\end{eqnarray}
It is also convenient to multiply (\ref{GroupCumulantsVsLNDistrContracted}) by $(-1)^{k} q^{2 k} / \lr{2 k}!$ and to sum over all $k$, which yields:
\begin{eqnarray}
\label{GroupPFVsLNW}
2/\pi S\lrs{C} \int \limits_{0}^{\pi} d\xi \: \sin^{2} \xi \: P\lrs{\xi; C} \cos\lr{q \xi/\pi }
 = W\lr{ q , -1}
\end{eqnarray}
Using the expression (\ref{PoissonSmearedConnectedGenFunc}) and performing the inverse Fourier transform w.r.t. the variable $q = \pi k$, the step distribution $P\lrs{g; C}$ can be finally found:
\begin{eqnarray}
\label{GroupProbDistr}
 \sin^{2}\xi \: P\lrs{\xi; C} = \eta \sum \limits_{k = - \infty}^{+ \infty}
\cos\lr{k \xi}   \lr{ \expa{-\pi^{2} \: k^{2} \epsilon\lrs{C} } \cos\lr{k \pi} - 1 }
\end{eqnarray}
\end{widetext}
where $\eta = \eta\lr{-1}$.

 For sufficiently small $\epsilon\lr{C}$ the function $\sin^{2}\xi \: P\lrs{\xi; C}$ (which is actually the probability distribution of steps over the coordinate $\xi$) looks like a gaussian peak near $\xi = \pi$. There is also a $\delta$-function singularity at $\xi=0$, which is irrelevant for diffusion. The width of the peak near $\xi = \pi$ is proportional to $ \sqrt{l_{vort} \: P\lrs{C}/S\lrs{C}}$. Thus assuming finite thickness of vortices smears the $\delta$-functions in (\ref{ScrDiffusionEquation}) and allows steps by any elements of the gauge group. For a fixed loop $C$ the width of the step distribution is proportional to vortex thickness, but if the vortex thickness $l_{vort}$ is kept fixed, for large loops deviations from equation (\ref{ScrDiffusionEquation}) tend to zero as $P\lrs{C}/S\lrs{C} \sim S\lrs{C}^{-1/2}$, thus asymptotically screening is recovered even for vortices of finite size, as it should be \cite{Greensite:07:1, DelDebbio:97}.

 The spectrum of string tensions which corresponds to the step distribution (\ref{GroupProbDistr}) can be found using the equation (\ref{StepDistrFromSpectrum}). As follows from (\ref{StepDistrFromSpectrum}), string tensions $\sigma_{R}$ are proportional to the expectation values of the characters $\chi_{R}\lr{g}$: $ \sigma_{R} = d_{R}^{-1} \int dg P\lrs{g; C} \chi_{R}\lr{g}$. Direct calculation gives to the first order in $\epsilon\lrs{C}$:
\begin{eqnarray}
\label{ThickVorticesStringTensions}
\sigma_{k} = 2 \eta -  2/3 \: \eta \: k\lr{k+1} \: \epsilon\lrs{C}, \quad k = 1/2, 3/2, \ldots
\nonumber \\
\sigma_{k} = 2/3 \: \eta \: k \lr{k + 1} \: \epsilon\lrs{C}, \quad k = 0, 1, 2, \ldots \quad
\end{eqnarray}
Remembering that $\epsilon\lrs{C} = l_{vort} P\lrs{C} / S\lrs{C}$, one can conclude that the difference between thin and thick vortices shows up only in perimeter-dependent effects, which can be taken into account if the step distribution $P\lrs{g; C}$ is allowed to depend on the loop $C$. Since perimeter dependence of Wilson loops corresponds to the contribution of charge self-energies, it follows from (\ref{ThickVorticesStringTensions}) that the energy required to screen a colour charge in $SU\lr{2}$ representation with spin $k$ is proportional to $k\lr{k + 1}$, i.e. to the second-order Casimir operator.

 It is interesting to note that Casimir scaling of string tensions can also be formally described using the expression (\ref{GroupPFVsLNW}), although the validity of such description should be discussed separately. In \cite{Greensite:07:1} it was shown that Casimir scaling at intermediate distances and asymptotic screening can be obtained for Yang-Mills vacuum dominated by center vortices if the field distribution $\xi^{a}\lr{z} f\lr{x}$ inside each center vortex is random at some scale $l_{c} \ll l_{vort}$. In this case the distribution of $\tilde{\mathbb{L}}\lrs{C; \Sigma_{z}}$ should be completely different from the distributions (\ref{PoissonDiffDistr}) and (\ref{PoissonDiffDistrSmeared}) if the size of the loop is comparable with the vortex thickness $l_{vort}$. By virtue of the Central Limit Theorem it should look like a gaussian distribution with the dispersion $\delta f^{2} l_{c}^{2} S\lrs{C}$, where $\delta f^{2}$ is the dispersion of $f\lr{x}$. For such distribution $W\lr{q} = - \delta f^{2} \: l_{c}^{2} \: S\lrs{C} q^{2}$, and the equation (\ref{GroupPFVsLNW}) yields the following result:
\begin{eqnarray}
\label{GroupProbDistrCS}
\sin^{2}\xi \: P\lrs{\xi; C}
 =
- \pi^{2} \delta f^{2} \: l_{c}^{2}  \sum \limits_{k = - \infty}^{+ \infty}
 k^{2} \: \cos\lr{k \xi}
= \nonumber \\ =
\pi \delta f^{2} \: l_{c}^{2} \: \frac{d^{2} }{d\xi^{2}} \: \delta\lr{\xi}
\end{eqnarray}
The distribution described by (\ref{GroupProbDistrCS}) is singular, but the string tensions $\sigma_{k}$ can nevertheless be calculated using the equation (\ref{StepDistrFromSpectrum}):
\begin{eqnarray}
\label{CharactersCS}
\sigma_{k} = - \frac{2}{\pi \: d_{k}} \: \int \limits_{0}^{\pi} d\xi \: \sin^{2}\xi \: P\lrs{\xi; C} \chi_{k}\lr{\xi}
= \nonumber \\ =
- \frac{2 \delta f^{2} \: l_{c}^{2}}{2 k + 1} \: \lim \limits_{\xi \rightarrow 0} \frac{d^{2}}{d \xi^{2}} \frac{\sin\lr{2 k + 1} \xi}{\sin \xi}
= \nonumber \\ =
\frac{8 \: \delta f^{2} \: l_{c}^{2}}{3} \: k \lr{k+1} \quad
\end{eqnarray}
 Thus such rather general assumptions concerning the distribution of the modified linking number $\tilde{\mathbb{L}}\lrs{C; \Sigma_{z}}$ lead to Casimir scaling at intermediate distances and to screening at asymptotically large distances, in full accordance with the results obtained in \cite{Greensite:07:1}.

 Finally, it should be noted that as the step distribution (\ref{GroupProbDistr}) is nonzero on the whole group manifold, it still describes discontinuous changes of the holonomy $g\lrs{C}$. Step distribution which describes continuous changes of the holonomy $g\lrs{C}$ in nonsingular gauge fields can only contain singularities of the form $\nabla_{a_{1}} \ldots \nabla_{a_{k}} \delta\lr{g,1}$ with finite $k$, as in (\ref{GroupProbDistrCS}). In order to obtain such step distribution for the model of thick center vortices, some more advanced assumptions on the distribution of linking numbers and on the distribution of fields inside vortices should be made. In particular, one could take into account that interactions between vortices should lead to deviations from Poisson-like distribution of the linking number. Although it could be extremely interesting to investigate such effects, this is not the aim of this paper. The derivation presented above is only intended to illustrate how the equation (\ref{ScrDiffusionEquation}) can arise as the limiting case of some more general diffusion equation and how the spectrum of string tensions can change in this case.

\section{Conclusions}
\label{sec:Conclusions}

\begin{figure}
  \includegraphics[width=5cm, angle=-90]{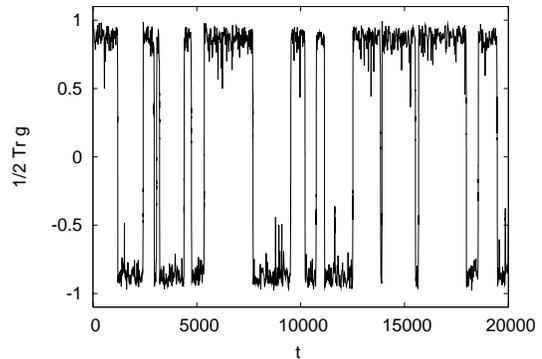}\\
  \caption{Time dependence of $1/2 \: \tr g$ for a particular implementation of a random walk on $SU\lr{2}$ group manifold with external potential $V\lr{g} = - \tr g^{2}$.}
  \label{fig:implementation}
\end{figure}

 In this paper a phenomenological analysis of \cite{Brzoska:05, Arcioni:05, Buividovich:06:2} was extended to the case of a general free random walk on the group manifold basing on the exact results obtained in \cite{Buividovich:06:3}. It was shown that the distribution of steps of a random walk of the holonomy $g\lrs{C}$ is directly related to the spectrum of string tensions of the theory and also to certain vacuum expectation values of the curvature tensor of Yang-Mills fields. In some sense the spectrum of string tensions of the theory is a dual image of the spatial distribution of non-Abelian magnetic flux, just as the amplitude of particle scattering in an external potential is proportional to the Fourier transform of this potential. The description of the dynamics of Wilson loops in terms of random walks on the gauge group manifold makes this correspondence more apparent.

 The relation between the step distribution of the random walk of $g\lrs{C}$ and the spectrum of string tensions was  used to analyze the random walk of Wilson loops in the case of complete screening, when string tension $\sigma_{R}$ depends only on the $N$-ality of the representation $R$. It turned out that when colour charges are completely screened, the holonomy $g\lrs{C}$ can only change by the elements of the group center, which implies that the only field configurations which contribute to the cumulants of the shifted curvature tensor are thin center vortices \cite{tHooft:78}. Complete screening, however, is only an asymptotic property, and the equation (\ref{ScrDiffusionEquation}) can as well be the $S\lrs{C} \rightarrow \infty$ limit of some more general diffusion equation. This possibility was investigated for the case of thick center vortices in Yang-Mills theory with $SU\lr{2}$ gauge group. It was shown that indeed for loops of finite size the equation (\ref{ScrDiffusionEquation}) can only be valid if center vortices are infinitely thin, but in the limit $S\lrs{C} \rightarrow \infty$ it is recovered even for vortices of finite thickness. For sufficiently small vortex sizes or for sufficiently large loops the distribution of steps $P\lrs{g; C}$ is a gaussian-like peak near $g = -1$ with the width $\epsilon\lrs{C} = l_{vort} P\lrs{C} / S\lrs{C}$. Of course these conclusions are model-dependent, but at least qualitatively the emergence of jumps by the elements of the group center can be understood. It could be also interesting to study how the equation (\ref{ScrDiffusionEquation}) is modified due to contribution of other field configurations such as instantons or monopoles.

 It seems that the most important property of random walks on Lie groups which lead to asymptotical screening-like effects is the possibility of fast jump-like transitions between points on the group manifold which differ by the elements of the group center. For instance, if the free diffusion equation (\ref{FreeDiffusionEquation}) is modified by introducing external $Z_{N}$-symmetric potential $V\lr{g}$, as proposed in \cite{Brzoska:05, Arcioni:05}, asymptotically stable probability distribution $p\lr{g} \sim \expa{ - V\lr{g}}$ is also $Z_{N}$-symmetric and has $N$ equal maxima at some points $\{ g z \}$, $z \in Z_{N}$. Random walker spends most time in the vicinity of such points and can only travel from one such point to another in a relatively short time. This effect is somewhat similar to the restoration of classically broken symmetries due to tunneling. In order to illustrate this example, random walk on $SU\lr{2}$ group with the external potential $V\lr{g} = - \tr g^{2}$ was simulated.  Time dependence of $1/2 \tr g$ for a particular implementation of such random walk is plotted on Fig. \ref{fig:implementation}. It can be seen that indeed the random walker mostly wanders near the points $g = \pm 1$, while the transitions between the regions close to these points are very fast and rare. It is reasonable to conjecture that such transitions simulate the discrete jumps described by the equation (\ref{ScrDiffusionEquation}) and lead to asymptotic screening \cite{Brzoska:05, Arcioni:05}.

 It could be very interesting to investigate whether such jumps could emerge in some explicitly $SU\lr{N}$-symmetric dynamical system. It could be also useful to consider some generalizations of a free random walk described by the equation (\ref{DiffEqIntegralGeneral}). For instance, one could consider random walks with memory, which can be more appropriate for the description of domain-like structure of Yang-Mills vacuum \cite{Greensite:07:1, Dosch:02, Ambjorn:80, Nielsen:79:1}.

\begin{acknowledgments}
 P. V. Buividovich is grateful to all members of the ITEP lattice group (ITEP, Moscow) for support and stimulating discussions, and especially to E. Luschevskaya for her kind hospitality. Illuminating discussions with V. I. Zakharov are also acknowledged. M. I. Polikarpov was partially supported by grants RFBR-05-02-16306a, RFBR-0402-16079a, RFBR-0602-04010-NNIOa and EU Integrated Infrastructure Initiative Hadron Physics (I3HP) under contract RII3-CT-2004-506078. P. V. Buividovich was partially supported by the grant of ICTP office of external activities for his participation in the 7th international conference "Symmetry in nonlinear mathematical physics" (Kyiv, June 24 -- 30, 2007), where a part of this work was reported.
\end{acknowledgments}


\end{document}